\documentclass[aps,prb,twocolumn]{revtex4}
\usepackage{amsmath,graphicx}

\begin{document}

\title{Scaling Properties of 1D Anderson Model with Correlated Diagonal
Disorder}
\author{L. I. Deych}
\author{M. V. Erementchouk}
\author{A. A. Lisyansky}

\begin{abstract}
Statistical and scaling properties of the Lyapunov exponent for a
tight-binding model with the diagonal disorder described by a dichotomic
process are considered near the band edge. The effect of correlations on
scaling properties is discussed. It is shown that correlations lead to an
additional parameter governing the validity of single parameter scaling.
\end{abstract}

\address{Physics Department, Queens College of the City University of New
York, Flushing, NY 11367}

\pacs{72.15.Rn,05.40.-a,42.25.Dd,71.55.Jv}
\maketitle

\section{Introduction}

It has been known for almost forty years that in standard one-dimensional
disordered models all states are localized for any strength of disorder, and
that there is no localization transition in such systems.\cite{gang,LGP}
Formally this is expressed by the statement that the Lyapunov exponent (LE),
$\gamma $, defined as
\begin{equation}
\gamma =-\lim_{L\rightarrow \infty }\frac{1}{2L}\ln {T},  \label{LE}
\end{equation}%
where $T$ is the transmission coefficient through the system of length $L $,
is always positive.\cite{LGP} While it may seem that there is nothing more
to be said about one-dimensional models, their localization properties have
recently attracted a great deal of attention. This renewed interest has been
concentrated in two areas. The first area includes studies of unusual models
that demonstrate the presence of extended states. It was noted first in Ref.~%
\onlinecite{dimer} that introducing correlations in the statistical
properties of random site energies in the Anderson model, one can create
extended states in the one-dimensional Anderson model at certain values of
energy. Later on, the authors of Ref.~\onlinecite{Moura} showed that regions
of extended states exist in 1D systems with long range correlations of the
random potential. The authors of Ref.~\onlinecite{Izrailev} demonstrated
that vanishing LE can be obtained in an arbitrarily chosen spectral region
by selecting a special form of the correlation function of the random
potential. It should be noted, however, that this result was obtained in
Ref.~\onlinecite{Izrailev} only in the second order of the weak disorder
expansion. Taking into account the next terms of the expansion makes LE
positive again.\cite{fourth order pt}

The second area of research interest in the field of one-dimensional
localization is focused upon scaling and statistical properties of finite
size systems. The single parameter scaling put forward in Ref.~%
\onlinecite{gang} is a cornerstone of the current approach to the Anderson
localization, but its interpretation in the presence of non-self-averaging
fluctuations of conductivity was the subject of long debate. Eventually it
was understood that the SPS hypothesis in Anderson localization means that
the entire distribution function of conductivity (or transmission), and not
individual moments, must be parameterized by a single parameter.\cite%
{Anderson,Shapiro} This distribution in one-dimensional systems has been
intensively studied\cite{KramerReview}, and it was established that the bulk
of the distribution for sufficiently long systems has a log-normal form. It
was also understood starting with Ref.~\onlinecite{Anderson} that under
certain circumstances the two parameters of the distribution, the average
value of the finite size LE, $\tilde{\gamma}=-\ln {T}/(2L)$, which coincides
with its limiting value, $\gamma =\langle \tilde{\gamma}\rangle $, and the
variance, $\sigma ^{2}=var(\tilde{\gamma}),$ are related to each other in a
universal way
\begin{equation}
\tau =\frac{\sigma ^{2}L}{\gamma }=1.  \label{SPS}
\end{equation}%
This relation reduces two parameters of the distribution to only one, and
provides, therefore, a justification and interpretation for SPS. Conditions,
under which Eq.~(\ref{SPS}) holds, however, have been established only
recently, when the authors of Ref.~\onlinecite{PRLclassics} showed that the
validity of SPS is controlled by a new macroscopic length, $l_{s}$, defined
in terms of the integral density of states, $N(E)$:
\begin{equation}
l_{s}^{-1}=\sin {(\pi N(E))}.  \label{eq:ls_definition}
\end{equation}%
where $E$ is energy. It was shown that SPS holds when $\kappa =(\gamma
l_{s})^{-1}>1$, and fails in the opposite case. Recently, the authors of the
present paper showed that parameter $\kappa $ plays a more important role
than just establishing the criterion for SPS. It was found that in the
non-SPS region the function $\tau $ defined in Eq.~(\ref{SPS}), and the
similar dimensionless combination of the third moment with the length $L$
and the Lyapunov exponent $\gamma $ are functions of this single parameter.%
\cite{nonSPSlast} In the recent paper Ref.~\onlinecite{Titov}, higher
moments of the distribution of the Lyapunov exponent were calculated
analytically for a quantum particle in a potential described by the Gaussian
delta-correlated random function. It was found that all moments of the
distribution for this model can be expressed as functions of the
localization length and the parameter $E/D^{2/3}$ , where $D$ characterizes
the strength of the potential. For the Gaussian white noise potential the
parameter $\kappa $ is also a function of $E/D^{2/3},$\cite{PRLclassics}
thus in the particular case of the Gaussian white noise potential these two
parameters are equivalent.

In the present paper, we combine the two discussed areas and focus upon
correlation induced changes in the probability distribution function of the
Lyapunov exponent. In order to study this problem, we consider a
tight-binding model with the random potential described by a zero-mean
dichotomic process. This process is characterized by an exponential
correlation function with a correlation radius, which is a free parameter of
the model. There is no localization-delocalization transition in this model,
and we are interested in effects of the correlation radius on scaling and
statistical properties of conductance in the regime of strongly localized
states. This model allows for an approximate analytical treatment in the
spectral region, where the correlation radius becomes a dominant length.
Combining analytical and numerical calculations we show that the presence of
this additional length significantly changes the transition between SPS and
non-SPS regions, and that in the non-SPS region it results in a new scaling
behavior.

\section{Quasiclassical approximation for Lyapunov exponent}

We consider the tight binding model, described by the equation of motion
\begin{equation}
\psi _{n+1}+\psi _{n-1}+(V\zeta _{n}-E)\psi _{n}=0,  \label{eq:eq_of_motion}
\end{equation}%
where $V$ describes the strength of the random potential. The potential
randomly takes one of two values $\pm V$ depending upon whether the
dichotomic random variable $\zeta _{n}$ is equal to $1$ or $-1$. The assumed
value remains constant within regions of random lengths (segments), which
are distributed according to the discrete analog of the Poisson
distribution. The average length of such regions is
\begin{equation}
l_{c}=\coth \left( \frac{1}{2r_{c}}\right) ,  \label{eq:average_length}
\end{equation}%
where $r_{c}$ is the correlation radius defined by
\begin{equation}
\left\langle \zeta _{n}\zeta _{m}\right\rangle =e^{-|n-m|/r_{c}}.
\label{eq:corr_definition}
\end{equation}%
The finite size LE is defined through the norm of the transfer matrix:
\begin{equation}
\tilde{\gamma}(E)=\frac{1}{L}\log \left\Vert T_{N}\ldots T_{1}\right\Vert ,
\label{eq:LE_def}
\end{equation}%
where $T_{n}$ are transfer matrices which, in our case, are equal to either $%
T_{+}$, or $T_{-}$ depending upon the value of $\zeta _{n}$, where $T_{\pm }$
are
\begin{equation}
T_{\pm }=\left(
\begin{array}{cc}
E\pm V & -1 \\
1 & 0\
\end{array}%
\right) .  \label{eq:transfer_mat_def}
\end{equation}

To consider the spectrum of this system let us note that an addition of a
constant potential $\pm U$ shifts a conduction band, $-2\leq E\leq 2$, of
the homogeneous tight-binding system with no on-site potential by the value
of the potential: $-2\mp U\leq E\leq 2\mp U$. Correspondingly, the spectrum
of the random system under consideration can be divided into three regions
with the qualitatively different spectral and transport properties. One
region, $0<|E|<2-V$, corresponds to the common part of the conduction bands
of homogeneous systems with the site potentials equal to either $+V$ or $-V$%
. \ In the second spectral region, consisting of energies $2-V<E<2+V$ and $%
-2-V<E<-2+V$, extended states exist in a homogeneous system with the site
potentials equal to $-V$ or $+V$, respectively. Finally the third region, $%
2+V<|E|$ corresponds to the energies, where extended states would not arise
in homogeneous systems with either of the two possible values of the
potential. Obviously, no states can arise in this region also in the system
with the random distribution of the potential.

While all states in this model are localized, different spectral regions
still have different transport properties. Transport in the first region is
characterized by multiple scattering of propagating modes, and the
localization in this region results from the interference of the scattered
waves. Transport in the second region can be described as tunneling through
states, arising within the segments of the system, where the potential
produces potential wells between barriers formed by segments with the
opposite value of the potential. If the average length of the
\textquotedblleft barrier\textquotedblright\ regions is larger than the
penetration length under the barrier, the transport is mostly determined by
the under-barrier tunneling, and the interference effects related to phase
changes of the wave functions inside potential wells become less important.
We are mostly interested in this region, where correlations are expected to
result in the most non-trivial effects. One could expect that the transport
in this region can be described within the quasi-classical approximation,
and we will show below that this is indeed the case.

In the context of our problem, this approximation means neglecting
commutators between transfer matrices at different sites. Introducing LE in
a homogeneous system with the potential $U$ according to
\begin{equation}
\gamma _{0}(E;U)=\mathrm{Re}\left[ \lambda _{0}(E;U)\right] ,
\label{eq:LE_pure_def}
\end{equation}%
where
\begin{equation}
\lambda _{0}=\cosh ^{-1}\left( \frac{E-U}{2}\right) ,  \label{eq:lambda_def}
\end{equation}%
we can write a quasiclassical expression for LE for Eq.~(\ref%
{eq:eq_of_motion}) in the following form
\begin{equation}
\tilde{\gamma}(E)\approx \frac{1}{L}\sum_{n=1}^{N}\gamma _{0}(E;V\zeta _{n}).
\label{eq:LE_limit}
\end{equation}%
Using the fact that the $\zeta _{n}=\pm 1$, Eq.~(\ref{eq:LE_limit}) can be
rewritten in the form
\begin{equation}
\tilde{\gamma}=\gamma _{a}+\frac{\delta}{L}\sum_{n=1}^{N}\zeta _{n},
\label{eq:LE_final}
\end{equation}%
where
\begin{eqnarray}
\gamma _{a} &=&\frac{1}{2}\left[ \gamma _{0}(E;V)+\gamma _{0}(E;-V)\right] ,
\notag  \label{eq:LE_characteristics} \\
\delta &=&\frac{1}{2}\left[ \gamma _{0}(E;V)-\gamma _{0}(E;-V)\right].
\end{eqnarray}

The averaged finite-length LE, $\gamma =\left\langle \tilde{\gamma}%
\right\rangle $, is determined by $\gamma _{a}$, while its variance $\sigma
^{2}$ depends upon $\delta$:
\begin{equation*}
\sigma ^{2}=\left\langle \gamma ^{2}\right\rangle -\left\langle \gamma
\right\rangle ^{2}=\frac{\delta^2}{L}l_{c}.
\end{equation*}%
In the first spectral region this approximation yields vanishing LE, which
is quite understandable, because interference effects responsible for the
localization in this region are completely neglected.

Eq.~(\ref{eq:LE_limit}) can be formally obtained in the following way. The
relation between LE and the eigenvalues of the transfer matrix allows us to
write
\begin{equation}
2\cosh (\tilde{\gamma}L)=\mathrm{Tr}\,\left( T_{N,1}\right) .
\label{eq:LE_thru_Tr}
\end{equation}%
where the transfer matrix $T_{N,1}$ consists of sequences of $T_{\pm }$
\begin{equation}
T_{N,1}=T_{N}\ldots T_{1}=\ldots T_{+}^{n_{3}}T_{-}^{n_{2}}T_{+}^{n_{1}},
\label{eq:total_transfer}
\end{equation}%
Powers $n_{k}$ of transfer matrices $T_{\pm }$ represent the lengths of the
regions where the potential remains constant. The matrices $T_{\pm }$ can be
diagonalized $T_{\pm }^{n}=R_{\pm }S_{\pm }^{n}R_{\pm }^{-1}$, where
\begin{equation}
S_{\pm }^{n}= \left(
\begin{array}{cc}
e^{\lambda _{0}(\pm V)n} & 0 \\
0 & e^{-\lambda _{0}(\pm V)n}%
\end{array}%
\right)
\end{equation}%
Thus, introducing $\alpha _{12}=R_{+}^{-1}R_{-}$ and $\alpha
_{21}=R_{-}^{-1}R_{+}$, and noting that for long enough samples the effect
of matrices $R_{\pm }$ appearing at the end of the structure on LE is
negligible, we obtain that LE can be found from Eq. (\ref{eq:LE_thru_Tr})
with the transfer matrix replaced by
\begin{equation}
\tilde{T}_{N,1}=\ldots \alpha _{21}S_{+}^{n_{2}}\alpha _{12}S_{-}^{n_{1}}.
\label{eq:total_thru_S}
\end{equation}%
The last step in the procedure is to commute, for example, $S_{+}$ and $%
\alpha _{12}$, with the purpose of collecting all $S_{\pm }$ together. If
one begins this rearrangement from the right side of Eq. (\ref%
{eq:total_thru_S}), each of such permutations produces two terms. One of
them consists of the product of the partially ordered and the remaining
non-ordered parts, while the second one contains commutators of $S_{+}$ and $%
\alpha _{12}$. Repeating the procedure leads to the recurrent equation for
the part of the transfer matrix containing only products of commuting
diagonal matrices $S_{\pm}$
\begin{equation}
T_{N,1}=\mathcal{O}_{N,1}+\sum_{k\in \mathrm{even}}T_{N,p_{k+1}}\Gamma
_{n_{k}}\mathcal{O}_{p_{k-1},1},  \label{eq:self-consistent}
\end{equation}%
where we have used the fact that $\alpha _{12}\alpha _{21}=1$, and
introduced the commutator $\Gamma _{n}=\left[ S_{+}^{n},\alpha _{12}\right] $%
. Notation $\mathcal{O}_{p_{k-1},1}$ stays for the product of all $S_{\pm }$
between sites $1$, and $p_{k-1}$: $\mathcal{O}_{N,1}=\ldots
S_{+}^{n_{2}}S_{-}^{n_{1}}$, where $p_{k}=\sum_{l=1}^{k}n_{l}$. \ Retaining
the first term only and using Eq.~(\ref{eq:LE_thru_Tr}) we arrive at Eq.~(%
\ref{eq:LE_limit}).

Eq.~(\ref{eq:self-consistent}) can be used to obtain a representation for
the transfer matrix in terms of the polynomial in powers of the parameter $%
\delta$, which enters the expression for commutators $\Gamma _{n}$, and can
be considered as a small parameter in the case of weak disorder. However,
the resulting expression for the transfer matrix cannot be interpreted as a
weak disorder expansion for LE. The reason for this is the fact that the
power of $\delta$ is determined by the number of segments with different
values of the potential, and the whole polynomial is actually an expansion
in terms of the number of \textquotedblleft jumps\textquotedblright\ between
$V$ and $-V$. For instance, the term with $\delta^2$ results from
configurations with only one such jump, and there are only $N$ such
configurations. In the limit $N\rightarrow \infty $ they give zero
correction to the LE (in addition to the exponentially small probability of
such configurations). Thus, we can see that for long enough samples only
terms of the order of $\mathcal{N}\sim N/2$ make a significant contribution
to the LE because the number of such terms is proportional to the binomial
coefficient $C_{N}^{\mathcal{N}}$ and, therefore they give a linear in $N$
contribution to the $\log \left\Vert T\right\Vert $. The actual number of
configurations contributing to LE depends upon the average length of the
segments with constant potential; and the number of jumps, corresponding to
the optimal configurations can be estimated as $N/l_{c}$. It is clear,
therefore, that corrections to the first term of Eq.~(\ref%
{eq:self-consistent}) decrease with increasing $l_{c}$. We thus expect that
the approximation should work when $l_{loc}/l_{c}\ll 1$. It actually becomes
exact in the limit of infinite $r_{c}$ and $V.$\cite%
{Pastur_Figotin,LargeCoupling} This can be understood from the following
arguments. Deviations from Eq. (\ref{eq:LE_final}) are caused by the
interference of the scattered waves. But in the second spectral region such
scattered waves have to tunnel through the barriers, where the magnitude of
the scattered waves decreases and the interference is destroyed. Obviously,
the effectiveness of the destruction depends on the width of the potential
barriers, and the rate of decay of the wave functions in them. These
characteristics are determined by $l_{c}$ and the magnitude of the potential
respectively.

In the spectral region, $2-V<|E|<2+V$ only $\gamma _{0}(E;V)$ in
Eq.~(\ref{eq:LE_characteristics}) is different from zero, and we
have for $\sigma^2$
\begin{equation}
\sigma ^{2}L=\gamma ^{2}l_{c}.  \label{eq:sigma2}
\end{equation}%
This expression is obviously different from SPS Eq.~(\ref{SPS}), and is
valid when $\gamma l_{c}\gg 1$. In this case, scaling described by the
function $\tau (\kappa )$ is no longer valid, but Eq.~(\ref{eq:sigma2})
suggests that a new type of scaling appears. It can be conveniently
described with the help of a new scaling function, $S$,
\begin{equation}
S=\frac{\sigma ^{2}L}{\gamma ^{2}l_{c}}\sim 1.  \label{eq:new_scaling}
\end{equation}%
The transition between the two scaling regimes is controlled by
the parameter $\gamma l_{c}$. If the correlation radius is so
small that this parameter remains much less than unity for the
entire second spectral region, the regular scaling behavior
expressed by $\tau (\kappa )$ persists. The behavior expressed by
Eq.~(\ref{eq:new_scaling}) occurs when $\gamma l_{c}$ is greater
than unity for the entire region $2-V<|E|<2+V$. More detailed
analysis of the transition between the two types of behavior
requires a numerical approach, which we discuss in the next
section of the paper.

\begin{figure}[tbp]
\vspace{-0.1in} \hspace{-0.5in} 
\includegraphics[width=2.5 in,angle=-90]{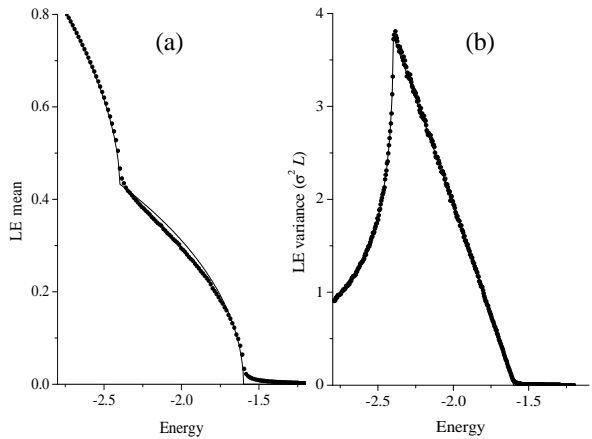}
\caption{Dependence of the mean of the Lyapunov exponent on
energy: numerical result (dotted line) and theory (solid line),
for potential $V=0.4$ and correlation radius $r_{c}=10$.}
\label{fig:LE}
\end{figure}

In the third spectral region, $|E|>2+V$, $S$ decreases with energy. This
behavior can be approximately described as
\begin{equation}
S\approx \frac{\left(\sqrt{2}\sqrt{|\epsilon |-1} - 1\right)^2}{\left( \sqrt{%
2}\sqrt{|\epsilon |-1} + 1\right) ^{2}}  \label{eq:S_third1}
\end{equation}%
for $0<|\epsilon |-1\ll 1$, where $\epsilon =(2-|E|)/U$ and
\begin{equation}
S\approx \left( \frac{1}{\epsilon }\right) ^{2}\frac{1}{(\log |U|+\log
|\epsilon |)^{2}}  \label{eq:S_third2}
\end{equation}%
for $|\epsilon |\gg 1$.

\section{Scaling properties of LE: numerical simulations}

For the numerical analysis, we calculate the LE iteratively using Eq.~(\ref%
{eq:LE_def}) and the usual technique of renormalization of the resultant
vector after every 10 iterations.\cite{Kramer} The length $l_{s}$ was
calculated according to Eq.~(\ref{eq:ls_definition}), where the integrated
density of states was obtained using the phase formalism \cite{LGP} with
consequent averaging over all realizations. To investigate scaling
properties we kept the length of chains much larger than\ the correlation
length, $l_{c},$ and the localization length, $l_{loc},$ for all magnitudes
of the random potential $U$ and values of the energy $E$ in the vicinity of
the second spectral region. Statistics were collected from $40000$
realizations. Figs.~1a and b present results of numerical computations of
the Lyapunov exponent. The numerical value of the correlation radius in
these figures is such that $l_{c}>l_{loc}$ for the entire $2-V<|E|<2+V$
spectral region. A comparison of numerical results and the approximation (%
\ref{eq:LE_limit}) is provided for both average LE and its variance. For
energies inside the first region this approximation gives zero, as discussed
above. In the second region, one can see that Eq.~(\ref{eq:LE_limit})
provides a reasonably good approximation for $\gamma $, and even better
agreement for $\sigma ^{2}$. In the third spectral region the agreement
between Eq.~(\ref{eq:LE_limit}) and the numerical calculations becomes
perfect.

\begin{figure}[tbp]
\vspace{-0.2in} \hspace{-0.5in} 
\includegraphics[width=2.5 in,angle=-90]{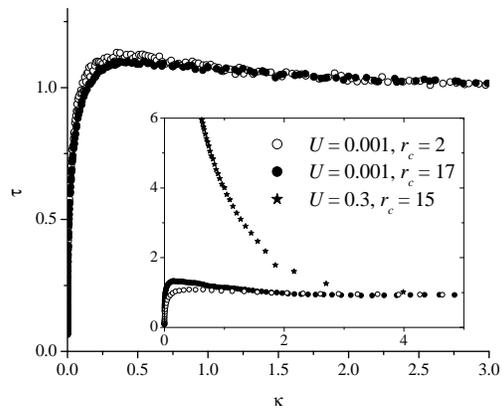}
\caption{$\protect\tau(\protect\kappa)$ for different
distributions of the potential: the uniform distribution (empty
circles) and the dichotomic process (filled circles) when $r_c$ is
the smallest length of the system.}
\label{fig:scalings_box_and_dich}
\end{figure}

We also studied numerically the scaling properties of our model.
Let us recall that for systems without correlations SPS is
controlled by the ratio $\kappa =l_{loc}/l_{s}$ of two macroscopic
lengths existing in the system. In the presence of the correlated
disorder the situation is more complicated because a new length
related to the correlation radius $r_{c}$ should be taken into
account. In our model this length is $l_{c}-$ the average length
of regions with constant potential. If $l_{c}$ is the smallest
length in the system, $l_{c}<(l_{loc},l_{s}),$ the presence of
correlations does not affect the scaling properties. Computer
simulations show that the scaling parameter $\tau $ behaves in
this case qualitatively similar to a system with uniformly
distributed non-correlated disorder. In
Fig.~\ref{fig:scalings_box_and_dich} the dependence of $\tau $ on
$\kappa $ is shown for both a tight-binding model with uniformly
distributed ($-U\leq V_{n}\leq U$) on-site potential (with $U$
ranging from $0.09$ to $0.145$) and the model under consideration.
For the latter, potentials ($V=0.001 - 0.004$) and correlation
radii ($r_{c}=2$ and $3$) where chosen such as to have $l_{c}$ the
smallest length up to the genuine spectral boundary.

\begin{figure}[tbp]
\vspace{-0.2in} \hspace{-0.5in} 
\includegraphics[width=2.5 in,angle=-90]{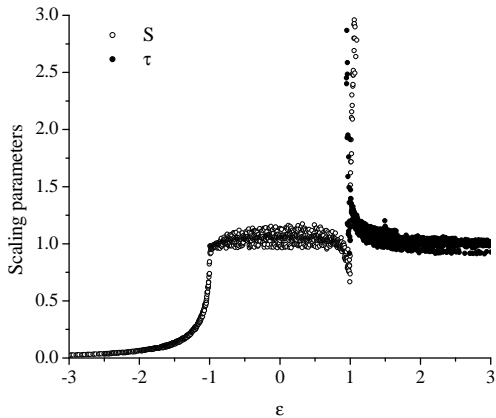}
\caption{The crossover between two scaling regimes, $\protect\tau
\sim 1$ (filled circles) and $S\sim 1$ (empty circles), as energy
variable $\protect \varepsilon =(E+2)/V$ changes, is shown for a
collection of different values of parameters ($V=0.2-1.2$ and
$r_{c}=5-250$) meeting the requirement $\protect\theta \gg 1.$}
\label{fig:crossover}
\end{figure}

However, when $l_{c}>l_{loc}$, the behavior of $\sigma ^{2}$
changes significantly, and the character of this change depends
upon the position of the spectral point, $E_{c},$ at which
$l_{c}=l_{loc}(E_{c})$. If $l_{c}$ is so large that $E_{c}$ falls
in the first spectral region $|E|<2-V$, the scaling function $\tau
$ remains essentially equal to unity up to the very small vicinity
of the boundaries of this interval. Inside a very small
neighborhood of points $|E|=2-V$ any scaling behavior disappears,
but immediately outside of this neighborhood inside the second
spectral interval $2-V<|E|<2+V$ a new scaling presented by
Eq.~(\ref{eq:new_scaling}) emerges. This situation is presented in
Fig.~\ref{fig:crossover}. In the third spectral region the
function $S$ decreases in agreement with the analytical formulas Eq.~(\ref%
{eq:S_third1}) and Eq.~(\ref{eq:S_third2}). It can be seen that the scaling
in terms of variable $\epsilon =(2-|E|)/U$ predicted by Eq.~(\ref%
{eq:S_third1}) persists over quite a wide interval of energies.

In the case when the energy $E_{c}$ belongs to the interior of the second
spectral region, the scaling properties are determined by the parameter $%
\kappa $. If $\kappa (E_{c})\ll 1$ we return to the situation of the
non-correlated disorder and if $\kappa (E_{c})>1$ the system does not
exhibit any scaling behavior in the second spectral region. The destruction
of scaling in this case may also be illustrated by the behavior of $\tau
(\kappa)$ for different values of $l_{c}$. When $l_{c}$ increases from the
values corresponding to the non-correlated limit, the function $\tau
(\kappa) $ starts changing as shown in the insert in Fig.~2 where the
dependence of $\tau $ is depicted for three different sets of parameters of
the random potential $U$ and $r_{c}$. It deviates from the universal (SPS)
dependence at larger values of $\kappa $ with the increasing magnitude of
the deviation depending upon both $r_{c}$ and $U$. It demonstrates in that
way the absence of scaling properties.

\section{Conclusion}

We considered statistical characteristics and scaling properties of the
Lyapunov exponent in a random medium with correlated disorder, using a
tight-binding model with a random potential described by a dichotomic
process. A simple theoretical approach for estimation of the statistics of
the Lyapunov exponent was suggested. Qualitative analysis based on this
model allowed us to estimate the effect of correlations on scaling
properties, and express this effect in terms of the ratio of the
localization length and the correlation length, $\theta =l_{c}/l_{loc}$.
When the correlation radius is much smaller than the localization length ($%
\theta \ll 1$), the system behaves as though correlations are absent and
demonstrates single parameter scaling, $\sigma ^{2}Ll_{loc}\sim 1,$ whose
validity is governed by the ratio $l_{loc}/l_{s}$. In the opposite case,
when $\theta \gg 1$, the crossover to another scaling, $\sigma
^{2}Ll_{loc}^{2}/l_{c}\sim 1,$ specific for the model under consideration,
takes place at values of energy close to the band gap.

\section*{Acknowledgments}

The authors would also like to thank Steve Schwarz for reading and
commenting on the manuscript. This work was supported by AFOSR
under Contract No. F49620-02-1-0305, PSC-CUNY grants, and
partially by NATO Linkage grant No. 978090.

\newpage

\end{document}